\documentclass[a4paper,11.5pt]{article}
\usepackage{blindtext}
\usepackage[a4paper, total={6.1in, 10in}]{geometry}
\usepackage{amsmath}
\usepackage{authblk}
\usepackage{subcaption}
\usepackage[final]{graphicx}
\usepackage{calc}
\usepackage{bm}
\usepackage{float}

\begin{document}

\title{Noise signal as input data in self-organized neural networks}
\author[1,2]{ V. Kagalovsky}
\author[1]{D. Nemirovsky} 
\author[3]{S.~V. Kravchenko}
\affil[1]{Shamoon College of Engineering, Beer-Sheva 84105, Israel}
\affil[2]{Max-Planck-Institut f\"ur Physik Komplexer Systeme, Dresden 01187, Germany}
\affil[3]{Physics Department, Northeastern University, Boston, Massachusetts 02115, USA}

\maketitle

\begin{abstract}

Self-organizing neural networks are used to analyze uncorrelated white noises of different distribution types (normal, triangular, and uniform).  The artificially generated noises are analyzed by clustering the measured time signal sequence samples without its preprocessing.  Using this approach, we analyze, for the first time, the current noise produced by a sliding  ``Wigner-crystal''-like structure in the insulating phase of a 2D electron system in silicon.  The possibilities of using the method for analyzing and comparing experimental data obtained by observing various effects in solid-state physics and simulated numerical data using theoretical models are discussed.

\end{abstract}

\section{Introduction}

The use of machine learning (ML) methods and, in particular, neural networks has become widespread in our time, especially in theoretical condensed matter physics~\cite{melko,oht1,oht2} based on numerical simulations producing state configurations sampled with Monte Carlo or wave functions.  On the other hand, intensive research is performed in the areas that require analysis of time series, such as speech recognition~\cite{Bianco2019}, geophysics~\cite{Thilo}, and radar technologies~\cite{Wan2019}.  Unfortunately, utilizing of neural networks for signal processing is limited, as a rule, by the suppression of unwanted noise and the localization of the useful signal patterns~\cite{Borodinov2019,Fotiadou2020,862121}.

\par However, not only the primary signal itself but also the accompanying noise sometimes carries valuable information about the behavior of the system under consideration.  Noise analysis is used, for example, in such dissimilar areas of science and technology as social media research~\cite{Woo2020} and nuclear reactor safety~\cite{Doney1994AcousticBD}.  

\par Successful attempts have also been made to classify artificially generated ``colored'' noises using both supervised machine learning~\cite{Nourbagheri2016} and self-organized neural networks methods~\cite{Bryant2014}.  In this article, we test the ability of the self-organizing map (SOM) approach to distinguish between subtypes of uncorrelated white noise that differ only in the shape of the distribution.  The main advantage of SOMs is that they consider both the distribution and the topology of the source data~\cite{misra2020}.

\par The main goal of this research is to analyze the current noise observed in a wide range of voltages (between two threshold voltages) when measuring voltage-current characteristics in the insulating state of a 2D electron system in silicon \cite{brussarski2018transport,shashkin2019recent,shashkin2021metal}.  Our paper is organized as follows.  In Section 2, we briefly describe the operations with SOM networks.  In Section 3, we apply SOM to learn various types of white noise.  Section 4 provides the experimental results, discusses the similarity of the two-threshold dependences observed in silicon and Type-II superconductors, and presents the SOM analysis of the experimental results.  In Section 5, we discuss our results and suggest future experimental and theoretical research for analyzing and comparing experimental data obtained by observing various effects in condensed matter physics.  Numerical data simulated using theoretical models are also discussed.

\section{SOM networks}
\par Although the foundations of the theory of self-organizing neural networks are described in many publications~\cite{Hykin,Zupan}, we will briefly recall the main provisions of the principle of their operation.  SOM networks are trained to distribute high dimensional input data (in our case, the sequence of random numbers) that have some common features between clusters formed by groups of neurons that make up the output layer.  This layer is represented, as a rule, by a two-dimensional rectangular or hexagonal lattice for greater clarity and ease of visualization~\cite{kohonen2001}.

\par
\begin{figure}
	\centering
	\scalebox{0.5}{\includegraphics[angle=0]{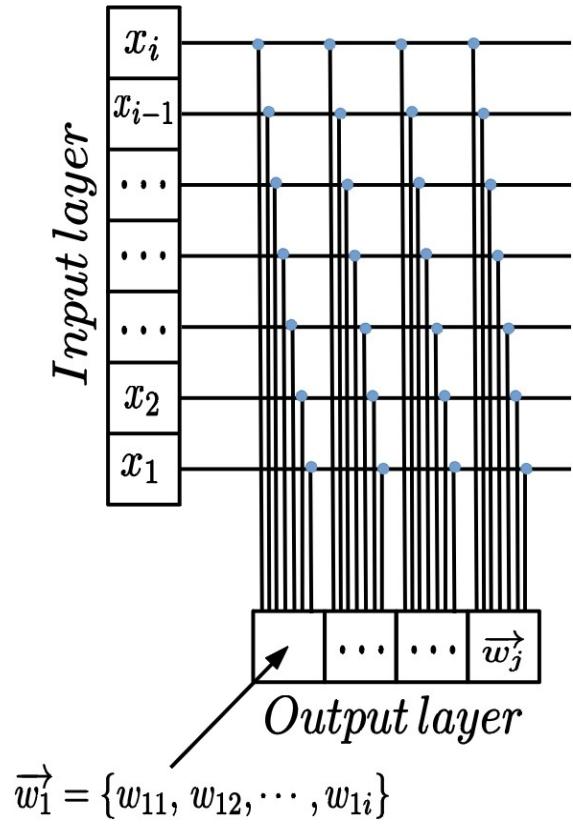}}
	\caption{SOM network structure consisting of i neurons of the input layer and j neurons of the output layer.}
	\label{fig1}
\end{figure}

 A typical SOM application procedure includes the following steps:

\begin{enumerate}
	\item  {\bf Input normalization}. Prior to passing data to the input layer, each of the elements of the input vector (sequence) is normalized in accordance with the following formula:
	\begin{equation}
		x_k=\frac{x_k}{\sqrt{\sum_{l=1}^i x^2_l }}
		\label{eq1}
	\end{equation}

where $x_k$ is the k-th element of the input vector whose total length is $i$ (see Fig. \ref{fig1}).

\item {\bf Training}. The weight vectors of the output neurons, $ \vec{w} $, are initialized using a random method.  Each of the $j$ weight vectors of the output layer neurons contains $i$ elements according to the number of input layer neurons (see Fig. \ref{fig1}).  After that, among all neurons of the output layer, a c-th neuron, for which the Euclidean distance to the input vector, $|| \vec{x}(N)-\vec{w}_c(N)||=\sum_{l=1}^{n} (w_{cl}(N) - x_{l}(N))^2$, is the smallest, is selected. Here $\vec{w}_c$ denotes the weight vector of the c-th neuron of the output level, and $N$ symbolizes a number of iterations correspondingly.  Thus, the neuron whose weight vector is closest to the input signal is declared the winner neuron.  At the next iteration step, the weight vector of the m-th output layer neuron is updated in accordance with the law:
\begin{eqnarray}
&\vec{w}_m(N+1) = \vec{w}_m(N)+ & \nonumber \\
& +\alpha (N) \cdot \exp \left( \frac{- || \vec{w}_c(N)-\vec{w}_m(N)||^2}{2 \sigma^2(N)} \right)\cdot (\vec{x}(N)-\vec{w}_m(N)) &
\label{update}
\end{eqnarray}
 where $\alpha(N)$ and $\sigma(N)$ are monotonically decreasing scalar functions of N. After making changes to the weight vectors of the output level (see Eq.~(\ref{update})), they should be normalized:
 
 \begin{equation}
 \sqrt{\sum_{l=1}^{n} w_{ml}^2(N)} = 1
 \end{equation}
 
  When the next data vector $\vec{x}(N+1)$ arrives at the input layer, the whole process repeats anew until convergence occurs (changes in $w$ are negligible) or the number of iterations exceeds a certain limit.
 
 \item {\bf Post-processing} Once the SOM training process is completed and the final form of the map is obtained, the input data is clustered, or other actions that are necessary to analyze the input data are performed~\cite{Vesanto1999}. 

\end{enumerate}

\section{Application of SOM to the analysis of various types of white noise}
\begin{figure}
	\begin{subfigure}{.33\textwidth}
		\centering
		\includegraphics[width=.99\linewidth]{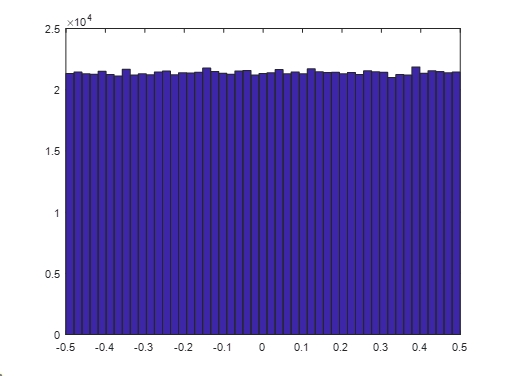}
		\caption{}
		\label{fig:sfig1}
	\end{subfigure}%
	\begin{subfigure}{.33\textwidth}
		\centering
		\includegraphics[width=.99\linewidth]{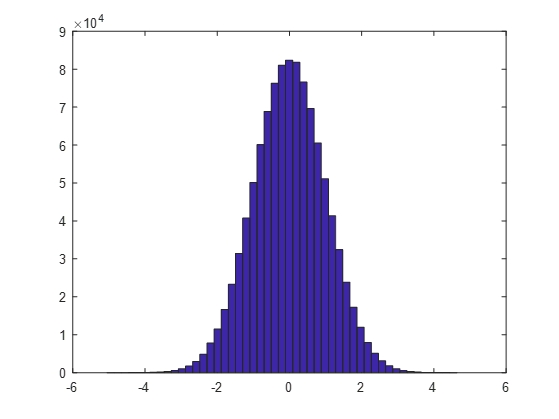}
		\caption{}
		\label{fig:sfig2}
	\end{subfigure}
	\begin{subfigure}{.33\textwidth}
	\centering
	\includegraphics[width=.99\linewidth]{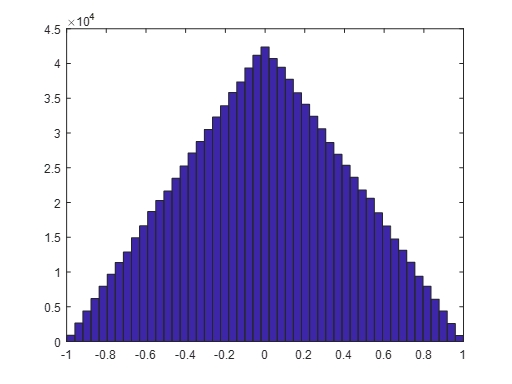}
	\caption{}
	\label{fig:sfig2}
\end{subfigure}
\caption{Histograms of different random signal distributions: (a) uniform, (b) normal and (c) triangular.}
\label{fig2}
\end{figure}

In this work, we analyze three different types of uncorrelated white noise, differing only in the shape of the distribution: uniform, normal and triangular.  Figure \ref{fig2} shows histograms of these distributions.

\begin{figure}
	\begin{subfigure}{.33\textwidth}
		\centering
		\includegraphics[width=.99\linewidth]{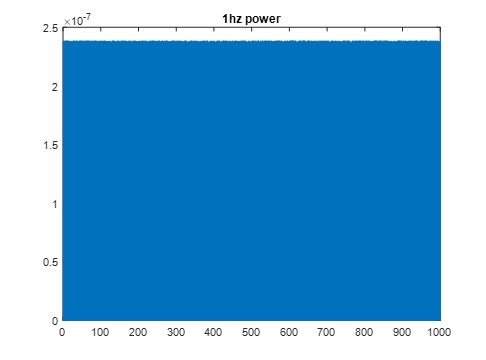}
		\caption{}
		\label{fig:sfig1}
	\end{subfigure}%
	\begin{subfigure}{.33\textwidth}
		\centering
		\includegraphics[width=.99\linewidth]{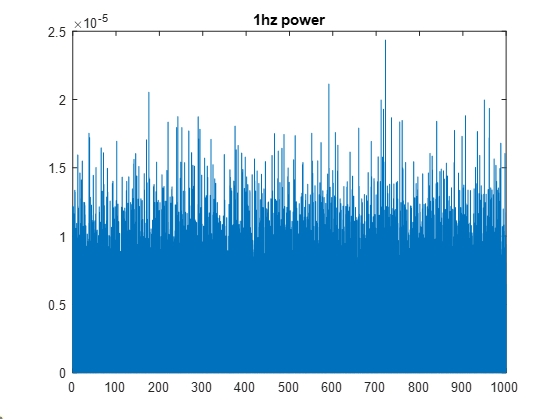}
		\caption{}
		\label{fig:sfig2}
	\end{subfigure}
	\begin{subfigure}{.33\textwidth}
		\centering
		\includegraphics[width=.99\linewidth]{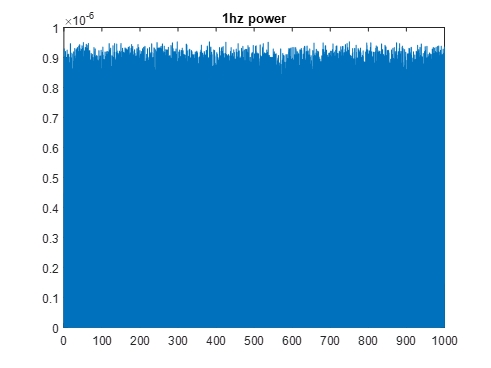}
		\caption{}
		\label{fig:sfig2}
	\end{subfigure}
	\caption{Power spectra of different random signal distributions: (a) uniform, (b) normal and (c) triangular.}
	\label{fig3}
\end{figure}

After generating the signals, the power spectra density of the random signals were analyzed to make sure that they are all varieties of white noise (see Fig. \ref{fig3}).  We also made sure that the autocorrelation function of each spectrum is the Dirac delta function.  Using the Matlab package, 512 signals of each kind of distribution were generated, each consisting of 1024 elements distributed accordingly.  After normalization (see Eq.~\ref{eq1}), a data array intended for training a neural network was formed by the concatenation of three sequentially (as suggested by~\cite{Bryant2014}) connected groups of signals: with uniform, normal and triangular distribution, containing 512 signals of each subspecies.  Thus, the data array for the system training process had dimensions of 1536 x 1024.

\begin{figure}
	\centering
	\scalebox{0.4}{\includegraphics[angle=0]{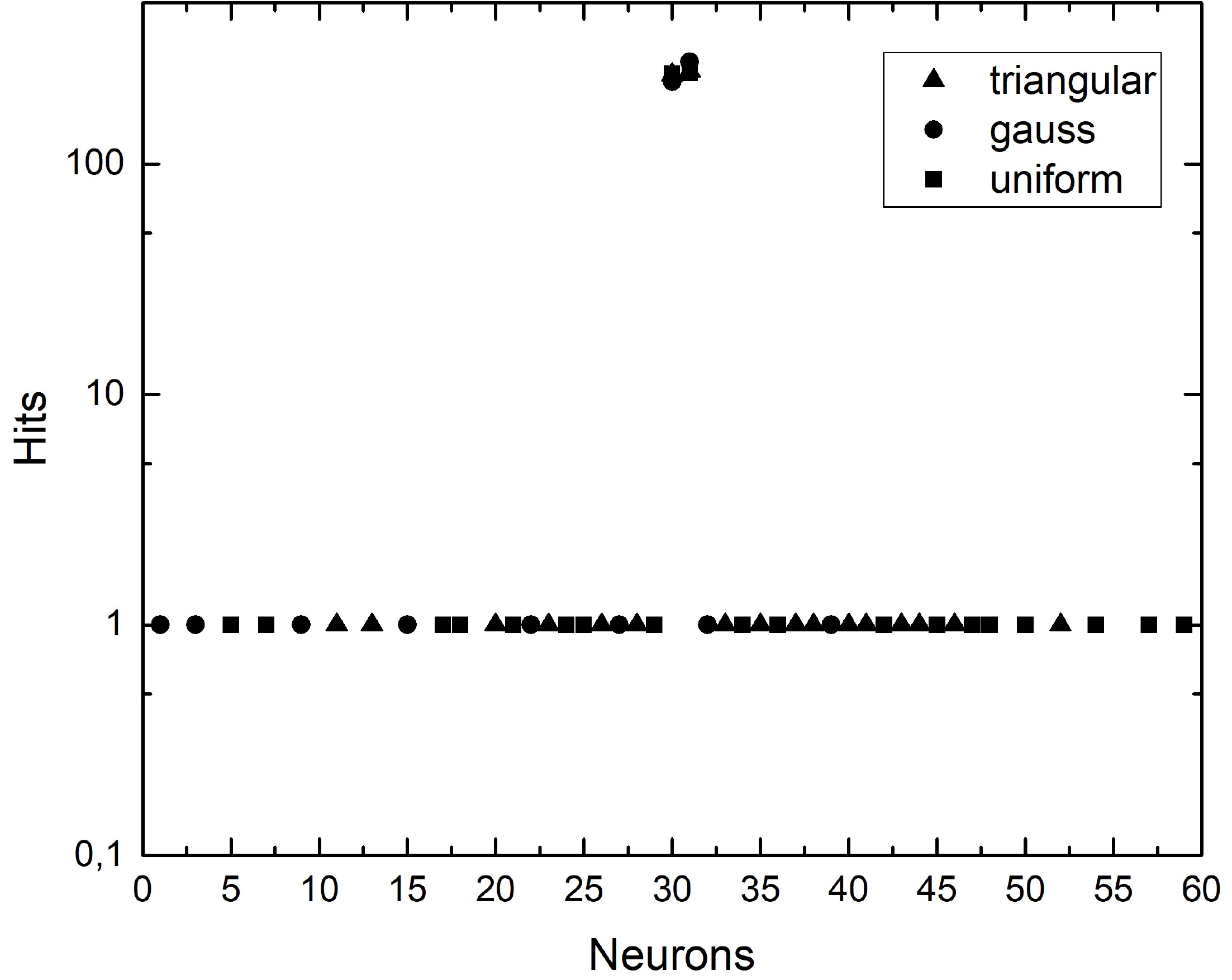}}
	\caption{Distribution of hits of training sets, related to different types of distribution, between neurons of the output layer of the trained SOM network. Neurons 60-64 are not shown as they were not hit at all during training.}
	\label{trained}
\end{figure}

Using the Matlab subroutine \texttt{selforgmap}, we generated a one-dimensional SOM network consisting of 64 neurons.  The SOM system was trained in batch unsupervised weight mode for 36000 epochs since the minimum recommended number of epochs~\cite{Hykin} should be at least 500 times the number of neurons in the output layer.  At the end of the SOM network training process, each of the 512 x 3 training sets is associated with one of the output layer neurons, as shown in Figure~\ref{trained}.  An interesting feature is observed in the distribution of training sets between output neurons, which will subsequently help us to distinguish between mixed signals: despite the fact that the vast majority of sets for all types of distributions fall exclusively on neurons 30 and 31 (46.6\% and 50.7\%, respectively), the remaining hits are distributed in such a way that the training set belonging to a certain type of distribution falls into only one group of neurons corresponding exclusively to this type of distribution (see Table~\ref{table1}).

\begin{figure}
	\begin{subfigure}{.33\textwidth}
		\centering
		\includegraphics[width=.99\linewidth]{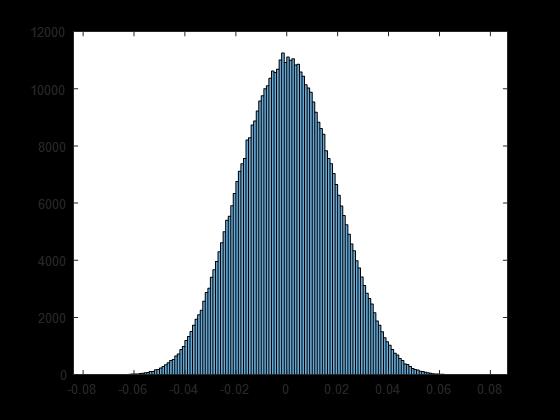}
		\caption{}
		\label{fig:sfig1}
	\end{subfigure}%
	\begin{subfigure}{.33\textwidth}
		\centering
		\includegraphics[width=.99\linewidth]{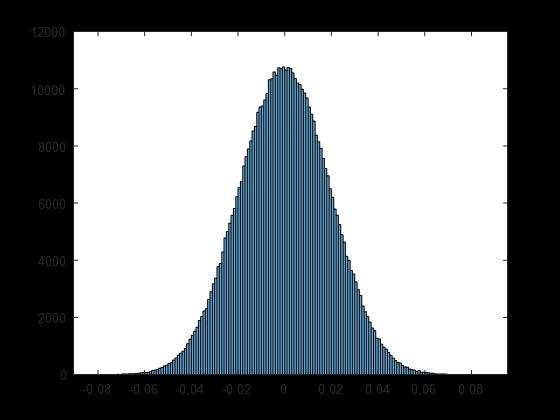}
		\caption{}
		\label{fig:sfig2}
	\end{subfigure}
	\begin{subfigure}{.33\textwidth}
		\centering
		\includegraphics[width=.99\linewidth]{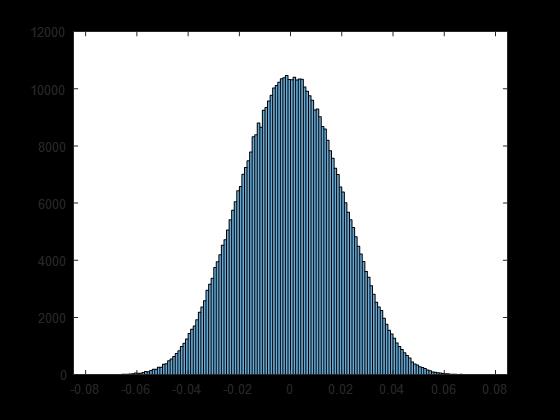}
		\caption{}
		\label{fig:sfig2}
	\end{subfigure}
	\caption{Histograms of different mixed distributions: (a) 50\% triangular + 30\% Gaussian + 20\% uniform, (b) 50\% Gaussian + 30\% uniform + 20\% triangular and (c) 50\% uniform + 30\% triangular + 20\% Gaussian, respectively. }
	\label{histo}
\end{figure}

\begin{table}
	
\begin{tabular}{|c|c|c|c|}
\hline
Distribution & Triangular & Gaussian & Uniform \\
\hline
Neurons & 11,13,20,23,28, & 1,3,9,15,22,27, & 7,17,18,21,24, \\
        & 33,35,37,38,40, & 32,39           & 25,29,34,36,42, \\
        & 41,43,44,46,52  &                 & 45,47,48,50,54, \\
        &                 &                 & 57,59      \\
\hline
  
\end{tabular}   
\caption{Non-overlapping groups of neurons corresponding to one of the distribution types.} 
\label{table1}
\end{table}

One of the tasks we set was to test the ability of the trained network to distinguish between data sets prepared from mixed types of distributions as follows.  Each data set related to a certain type of distribution was multiplied by a weight factor, after which it was added to another set related to a different type of distribution, also multiplied by the weight factor.  This procedure has been performed on each element of the set.  The following mixed data sets were prepared by this method: 50\% triangular + 30\% Gaussian + 20\% uniform, 50\% Gaussian + 30\% uniform + 20\% triangular, and 50\% uniform + 30\% triangular + 20\% Gaussian, respectively. 
\par
\begin{figure}
	\centering
	\scalebox{0.4}{\includegraphics[angle=0]{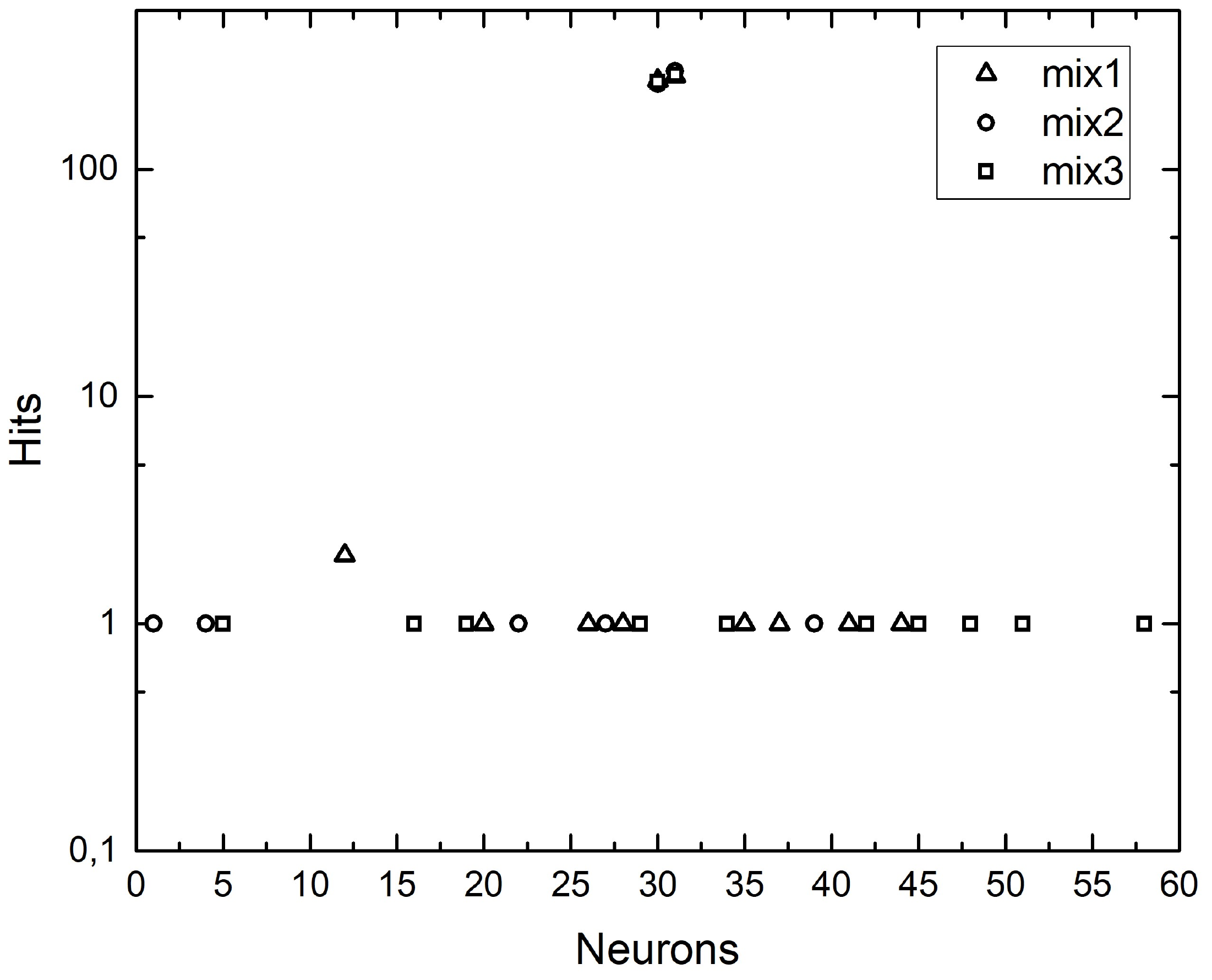}}
	\caption{Distribution of mixed sets between neurons of the output layer of the trained SOM network: mix 1 -- 50\% triangular + 30\% Gaussian + 20\% uniform (open triangles), mix2 -  50\% Gaussian + 30\% uniform + 20\% triangular (open circles) and mix3 -- 50\% uniform + 30\% triangular + 20\% Gaussian (open squares), respectively. Neurons 60-64 are not shown as they were not hit at all during testing.}
	\label{mixed}
\end{figure}

 Prior to analyzing the data sets using the trained network, we checked that the mixed sets have the properties of white noise (constant spectral density and no correlation) and also built histograms of the obtained distributions, shown in Figure~\ref{histo}.  The raw results of the analysis of mixed data sets are presented in Figure~\ref{mixed}.

\par For a quantitative analysis of the results obtained, it is necessary to introduce a similarity factor between the hit vectors of pure and mixed data sets.  The classical vector similarity criteria cannot give satisfactory results since the 30--th and 31--th elements of the hit vectors, consisting of 64 elements each, are almost the same for both pure and mixed sets and prevail over the rest of the hit vector elements.  In order to level the overwhelming values of 30--th and 31--th elements and take into account the impact of the remaining elements of the hit vector (which, in fact, make it possible to distinguish between different data sets), we introduced the following similarity criterion:

\begin{equation}
	\text{Similarity}(k,l)=\sum_{m=1}^{j=64} \text{sign}(hit_m^k*hit_m^l),
\end{equation}
where $hit_m^k$ is the m-th element of the k-th hit vector, which is the number of hits related to the k-th data set into the m-th neuron.

\begin{table}
	\begin{tabular}{|c|c|c|c|}
		\hline
		Distribution & Triangular & Gaussian & Uniform \\
		\hline
		50\% triangular &   &    & \\
		30\% Gaussian & 9  &  2  &  2\\
		20\% uniform &   &    & \\
		\hline
		50\% Gaussian &   &   & \\
		30\% uniform & 2 & 6 & 2 \\
		20\% triangular & & & \\
		\hline
		50\% uniform  & & & \\
		30\% triangular & 2 & 2 & 8\\
		20\% Gaussian & & & \\
		\hline
		
	\end{tabular}   
	\caption{Similarity factors between hit vectors of pure training sets and mixed sets.} 
		\label{table2}
\end{table}

\par The final test results are shown in Table~\ref{table2}.  From the results presented in Table 2, we can conclude that the trained network with 100\% accuracy determines the type of spectrum that makes the main contribution to the mixed data set.  However, it is still difficult to distinguish between secondary contributions.

\par It should be noted that this is our first attempt to analyze and compare various subtypes of white noise.  In the future, after obtaining a sufficient amount of experimental data, we will be able to analyze them in a similar way.

 \section{Experimental noise and its SOM analysis}
 \label{marker}
\begin{figure}
	\begin{center}
	\scalebox{0.52}{\includegraphics[angle=0]{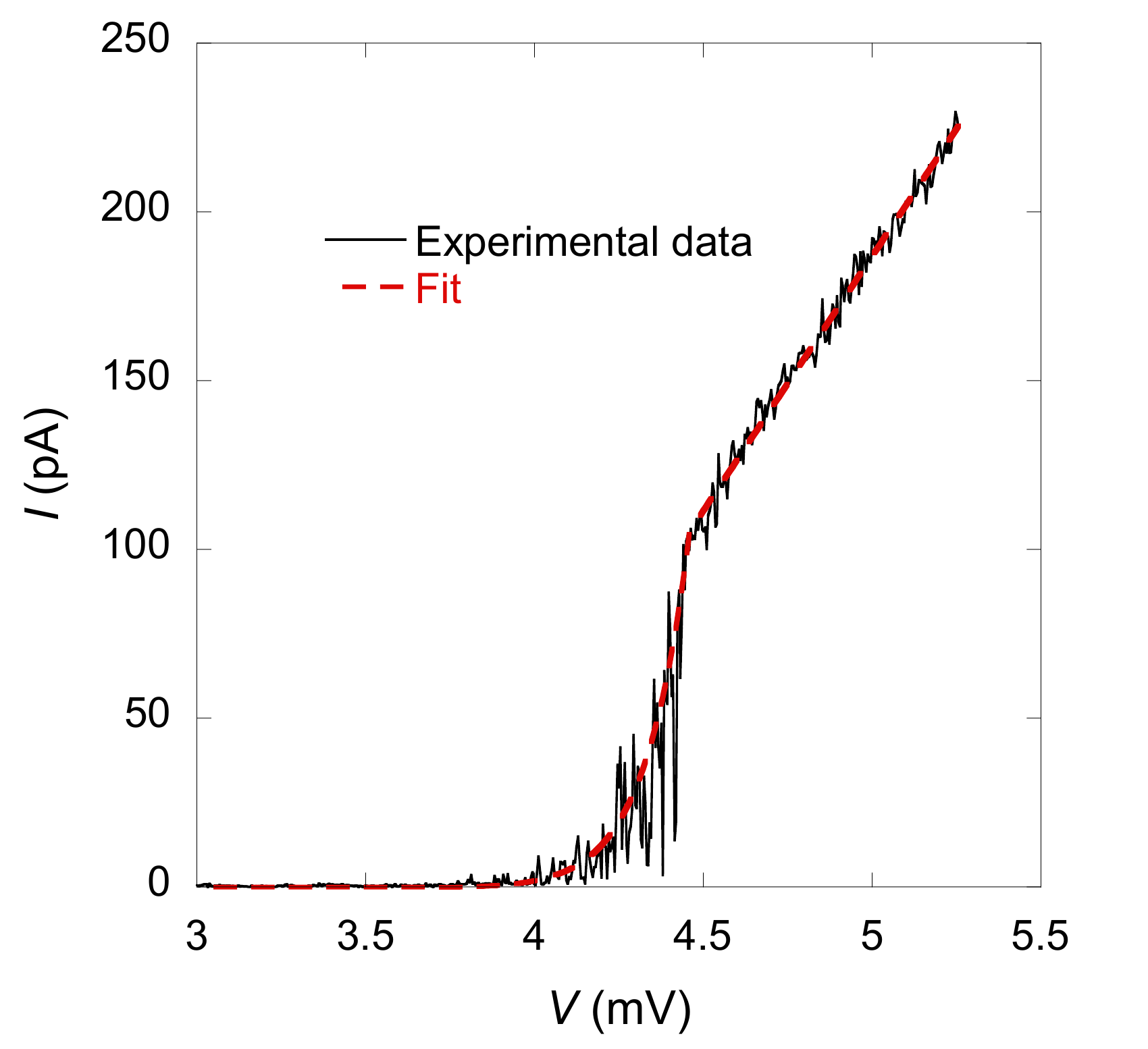}}
		\end{center}
\caption{Voltage-current characteristic for the electron density $n_{\text{s}}=5.36\times 10^{10}$~cm$^{-2}$ at a temperature of 60~mK. The dashed line is a fit to the data (for more on this, see Ref.~\cite{brussarski2018transport}).}

\label{v-i} 
\end{figure}

\begin{figure}
\begin{center}
	\scalebox{0.52}{\includegraphics[angle=0]{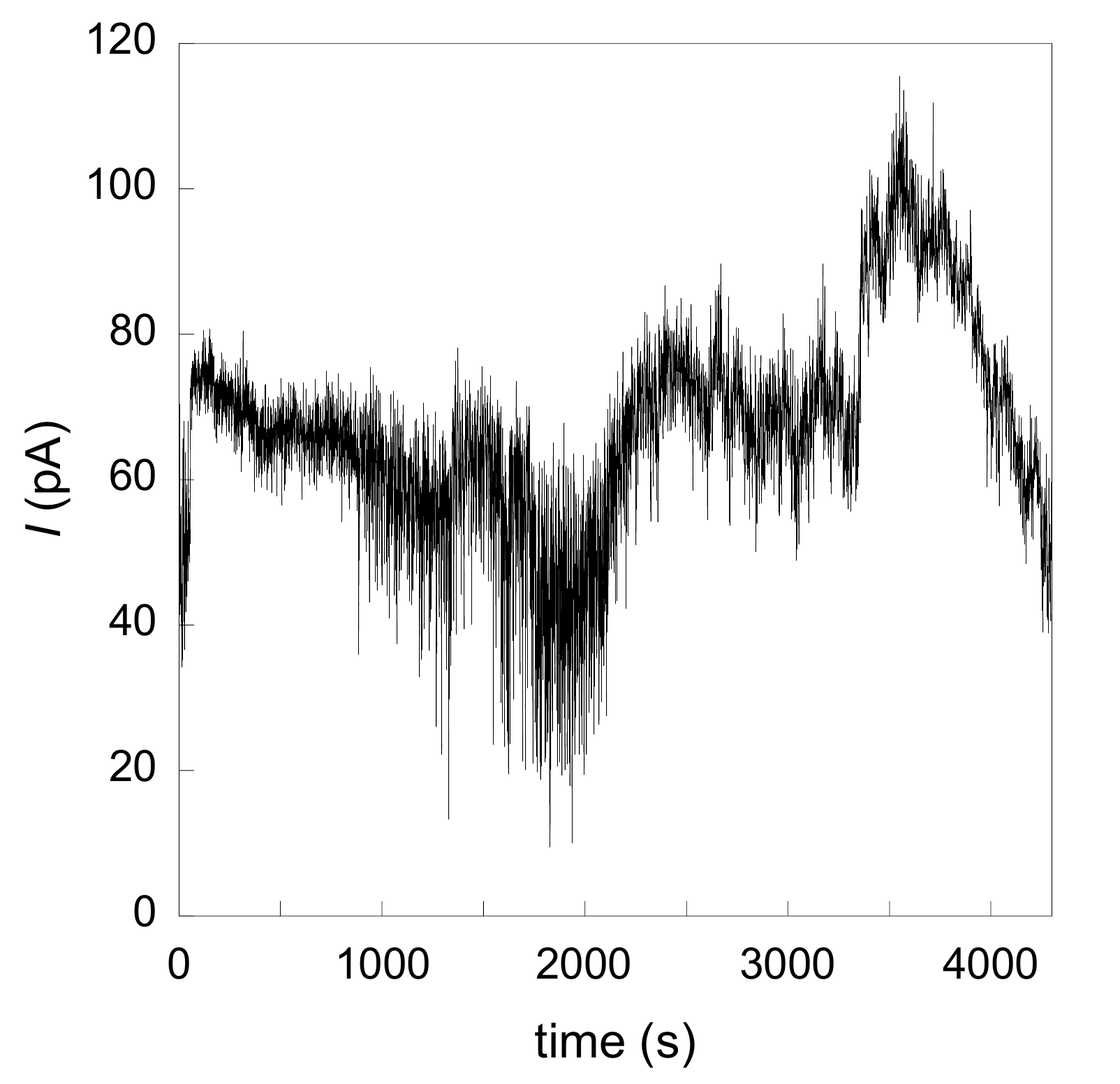}}
\end{center}
		\caption{The current as a function of time for $n_{\text{s}}=5.20\times 10^{10}$~$cm^{-2} $ and a temperature of 60~mK at a bias voltage of $V=4.90$~mV.
		 }
	\label{t-i} 
\end{figure}

Figure~\ref{v-i} shows a low-temperature voltage-current characteristic in the insulating regime, \textit{i.e.}, at electron densities below the metal-insulator transition. The samples studied are (100)-silicon MOSFETs with a peak electron mobility close to 3~m$^2$V$^{-1}$s$^{-1}$ at $T<0.1$~K (for information on the samples, see Ref.~\cite{heemskerk98}).  Voltage is applied between the source and the drain, and the induced current is measured.  The interaction parameter, \textit{i.e.}, the ratio of the Coulomb and Fermi energies, exceeds $\sim20$.  The observed features are indicative of the electron solid (deformed Wigner crystal) depinned by the applied bias voltage.
 The current is near zero at bias voltages up to approximately 4.0~mV.  Then, at bias voltages between approximately 4.0 and 4.4 mV, it sharply increases and exhibits strong fluctuations with time that are comparable to its value.  At yet higher bias voltages, the slope of the $V$-$I$ curve is significantly reduced, the noise becomes barely perceptible, and the $V$-$I$ curve becomes linear, although not ohmic.  Time fluctuations of the current between bias voltages 4.0-4.4~mV are shown in Fig.~\ref{t-i}.
 

\begin{figure}
\begin{center}
\scalebox{0.3}{\includegraphics[angle=0]{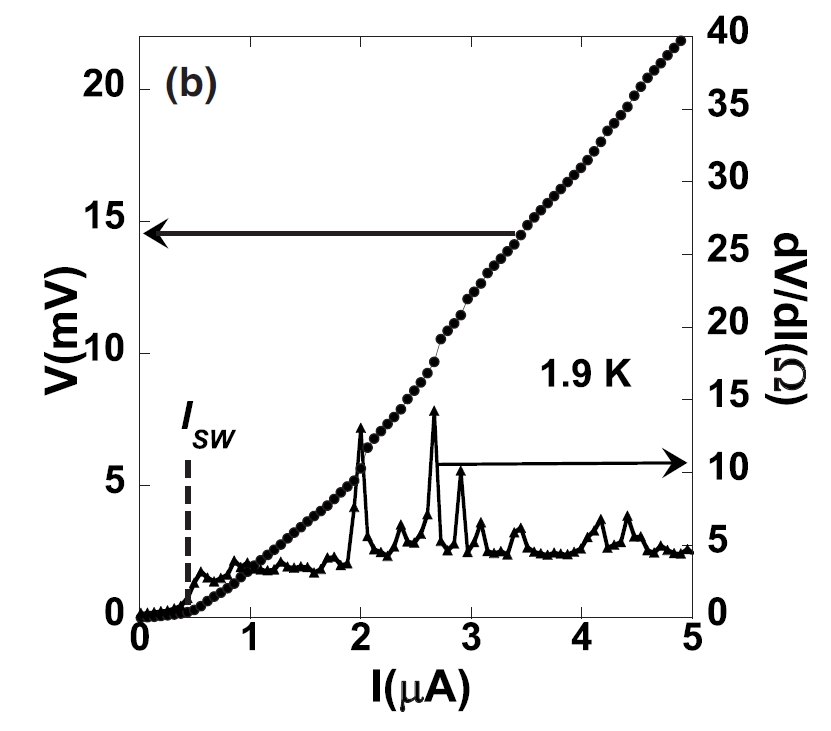}}
\end{center}
\caption{Voltage-current (V-I) curves and $dV/dI$ of the NbSe$_2$ NW device (adapted from~\cite{Res1}).}
\label{dvdi}
\end{figure}

There is a striking similarity between the double-threshold $V$-$I$ dependences in silicon MOSFET samples and those (with the voltage and current axes interchanged) in the type-II superconductors, where the existence of the vortex lattice has been firmly established~\cite{blatter94,Res1,Res2,Res3,Res4}.  An example of $I$-$V$ characteristic in such a system is shown in Fig.\ref{dvdi} adapted from Ref.~\cite{Res1}.  Voltage is zero at currents below approximately 0.4~$\mu$~A.  Then it increases nonlinearly, but above approximately 0.35~$\mu$~A, the $I$-$V$ dependence becomes linear.  In our case, the situation is reciprocal: a voltage is applied, but at first the current is zero in the limit of zero temperature; the depinning of the electron solid is signaled by the appearance of a non-zero current.  Unfortunately, Fig.\ref{dvdi} does not provide error bars for the voltage data, but the differential resistance exhibits strong fluctuations indicating strong noise signal.  The physics of the vortex lattice in Type-II superconductors can be successfully adapted for the case of an electron solid, as was shown by Valeri Dolgopolov \cite{dolgopolov2018similarity} (see also Ref.~\cite{brussarski2018transport}).

\par When preparing this article, no serious attempts were made to analyze the noise spectrum accompanying the measurements shown in Fig.~\ref{t-i} due to the lack of experimental data (signal length).  However, when tested by a trained neural network, the only signal with a length of 1024 elements fell on neuron 31 of the trained network, which suggests that it belongs to white noise.  Since this work represents only the initial stage of processing and analysis of the experimental data, in the future, we will be able to conduct a more thorough study of the noise that occurs during measurements.

\section {Conclusions and future research}
We have presented a Machine Learning procedure for self-organizing neural networks to study uncorrelated white noise and to analyze the noise measured in the experiment.  We have only recently embarked on this project and have not had enough experimental data (SOM study of the only experimental sample allows us to suggest that the data exhibits white noise).  But this initial approach allows us to suggest the future research necessary for analyzing and comparing experimental data obtained by observing various effects in condensed matter physics and numerical data, simulated using theoretical models.  Firstly, one of the authors (S.V.K.) will perform long enough measurements of the current as a function of time at different voltages between the two thresholds.  These results used as an input into the SOM network will allow us to achieve a more conclusive description of the noise measured in the experiment.  Secondly, we will try to obtain experimental data for noises measured in Type-II superconductors and compare two sets of experimental results to confirm a similarity between the two experiments.  Thirdly, the most challenging research direction will be to construct Hartree-Fock models to simulate currents in silicon system (we have already succeeded in building the Hartree-Fock description of the equilibrium system to reproduce the first threshold dependence on the temperature \cite{weHF}).  In parallel, we will simulate voltage noise produced by sliding Abrikosov lattice in Type-II superconductor when superconducting current exceeds the first threshold value, similar to numerical simulations of $I-V$ characteristics in a strong-pinning regime~\cite{Res4}.  We will then intercompare the results of these numerical simulations and compare them with the detailed experimental data.  

\section{Acknowledgements}
We greatly benefited from discussions with V. T. Dolgopolov.  With deep sadness, we inform the readers that Professor Dolgopolov recently passed away. 
V.K. and D.N. were supported by the Ukrainian-Israeli Scientific Research Program MESU and MOST grant 3-16430, and  by the SCE internal grant EXR01/Y17/T1/D3/Yr1.  S.V.K. was supported by NSF grant No.\ 1904024.

*\bibliography{references}

\begin{thebibliography}{10}
\expandafter\ifx\csname url\endcsname\relax
  \def\url#1{\texttt{#1}}\fi
\expandafter\ifx\csname urlprefix\endcsname\relax\def\urlprefix{URL }\fi
\providecommand{\bibinfo}[2]{#2}
\providecommand{\eprint}[2][]{\url{#2}}

\bibitem{melko}
\bibinfo{author}{Torlai, G.} \& \bibinfo{author}{Melko, R.~G.}
\newblock \bibinfo{title}{Learning thermodynamics with {B}oltzmann machines}.
\newblock \emph{\bibinfo{journal}{Phys. Rev. B}} \textbf{\bibinfo{volume}{94}},
  \bibinfo{pages}{165134} (\bibinfo{year}{2016}).

\bibitem{oht1}
\bibinfo{author}{Ohtsuki, T.} \& \bibinfo{author}{Ohtsuki, T.}
\newblock \bibinfo{title}{Deep learning the quantum phase transitions in random
  two-dimensional electron systems}.
\newblock \emph{\bibinfo{journal}{J. Phys. Soc. Japan}}
  \textbf{\bibinfo{volume}{85}}, \bibinfo{pages}{123706}
  (\bibinfo{year}{2016}).

\bibitem{oht2}
\bibinfo{author}{Ohtsuki, T.} \& \bibinfo{author}{Ohtsuki, T.}
\newblock \bibinfo{title}{Deep learning the quantum phase transitions in random
  electron systems: Applications to three dimensions}.
\newblock \emph{\bibinfo{journal}{J. Phys. Soc. Japan}}
  \textbf{\bibinfo{volume}{86}}, \bibinfo{pages}{044708}
  (\bibinfo{year}{2017}).

\bibitem{Bianco2019}
\bibinfo{author}{Bianco, M.~J.} \emph{et~al.}
\newblock \bibinfo{title}{{Machine learning in acoustics: Theory and
  applications}}.
\newblock \emph{\bibinfo{journal}{J. Acoust.}} \textbf{\bibinfo{volume}{146}},
  \bibinfo{pages}{3590--3628} (\bibinfo{year}{2019}).

\bibitem{Thilo}
\bibinfo{author}{Wrona, T.}, \bibinfo{author}{Pan, I.},
  \bibinfo{author}{Gawthorpe, R.~L.} \& \bibinfo{author}{Fossen, H.}
\newblock \bibinfo{title}{Seismic facies analysis using machine learning}.
\newblock \emph{\bibinfo{journal}{Geophysics}} \textbf{\bibinfo{volume}{83}},
  \bibinfo{pages}{O83--O95} (\bibinfo{year}{2018}).

\bibitem{Wan2019}
\bibinfo{author}{Wan, J.}, \bibinfo{author}{Chen, B.}, \bibinfo{author}{Xu,
  B.}, \bibinfo{author}{Liu, H.} \& \bibinfo{author}{Jin, L.}
\newblock \bibinfo{title}{{Convolutional neural networks for radar {HRRP}
  target recognition and rejection}}.
\newblock \emph{\bibinfo{journal}{EURASIP J. Adv. Signal Process.}}
  \textbf{\bibinfo{volume}{2019}}, \bibinfo{pages}{5} (\bibinfo{year}{2019}).

\bibitem{Borodinov2019}
\bibinfo{author}{Borodinov, N.} \emph{et~al.}
\newblock \bibinfo{title}{{Deep neural networks for understanding noisy data
  applied to physical property extraction in scanning probe microscopy}}.
\newblock \emph{\bibinfo{journal}{Npj Comput. Mater.}}
  \textbf{\bibinfo{volume}{5}}, \bibinfo{pages}{25} (\bibinfo{year}{2019}).

\bibitem{Fotiadou2020}
\bibinfo{author}{Fotiadou, E.} \& \bibinfo{author}{Vullings, R.}
\newblock \bibinfo{title}{Multi-channel fetal ecg denoising with deep
  convolutional neural networks}.
\newblock \emph{\bibinfo{journal}{Front. Pediatr.}}
  \textbf{\bibinfo{volume}{8}}, \bibinfo{pages}{508--508}
  (\bibinfo{year}{2020}).

\bibitem{862121}
\bibinfo{author}{Potamitis, I.}, \bibinfo{author}{Fakotakis, N.} \&
  \bibinfo{author}{Kokkinakis, G.}
\newblock \bibinfo{title}{Impulsive noise suppression using neural networks}.
\newblock In \emph{\bibinfo{booktitle}{2000 IEEE International Conference on
  Acoustics, Speech, and Signal Processing. Proceedings (Cat. No.00CH37100)}},
  vol.~\bibinfo{volume}{3}, \bibinfo{pages}{1871--1874} (\bibinfo{year}{2000}).

\bibitem{Woo2020}
\bibinfo{author}{Woo, J.} \& \bibinfo{author}{Yun, J.}
\newblock \bibinfo{title}{Content noise detection model using deep learning in
  web forums}.
\newblock \emph{\bibinfo{journal}{Sustainability}}
  \textbf{\bibinfo{volume}{12}}, \bibinfo{pages}{5074} (\bibinfo{year}{2020}).

\bibitem{Doney1994AcousticBD}
\bibinfo{author}{Doney, G.}
\newblock \bibinfo{title}{Acoustic boiling detection} (\bibinfo{year}{1994}).

\bibitem{Nourbagheri2016}
\bibinfo{author}{Nourbagheri, M.} \& \bibinfo{author}{Zohdy, M.}
\newblock \bibinfo{title}{Coloured noise signal identification using supervised
  learning algorithm}.
\newblock \emph{\bibinfo{journal}{Int. J. Comput. Sci. Inf. Technol. Res.}}
  \textbf{\bibinfo{volume}{5}}, \bibinfo{pages}{389--395}
  (\bibinfo{year}{2016}).

\bibitem{Bryant2014}
\bibinfo{author}{Bryant, T.} \& \bibinfo{author}{Zohdy, M.}
\newblock \bibinfo{title}{Noise signal identification by modified
  self-organizing maps}.
\newblock \emph{\bibinfo{journal}{Int. J. Comput. Sci. Inf. Technol. Res.}}
  \textbf{\bibinfo{volume}{03}}, \bibinfo{pages}{48--53}
  (\bibinfo{year}{2014}).

\bibitem{misra2020}
\bibinfo{author}{Siddharth~Misra, H.~L.} \& \bibinfo{author}{H, J.}
\newblock \emph{\bibinfo{title}{Machine Learning for Subsurface
  Characterization}} (\bibinfo{publisher}{Gulf Professional Publishing},
  \bibinfo{year}{2020}).

\bibitem{brussarski2018transport}
\bibinfo{author}{Brussarski, P.}, \bibinfo{author}{Li, S.},
  \bibinfo{author}{Kravchenko, S.~V.}, \bibinfo{author}{Shashkin, A.~A.} \&
  \bibinfo{author}{Sarachik, M.~P.}
\newblock \bibinfo{title}{Transport evidence for a sliding two-dimensional
  quantum electron solid}.
\newblock \emph{\bibinfo{journal}{Nat.\ Commun.}} \textbf{\bibinfo{volume}{9}},
  \bibinfo{pages}{3803} (\bibinfo{year}{2018}).

\bibitem{shashkin2019recent}
\bibinfo{author}{Shashkin, A.~A.} \& \bibinfo{author}{Kravchenko, S.~V.}
\newblock \bibinfo{title}{Recent developments in the field of the
  metal-insulator transition in two dimensions}.
\newblock \emph{\bibinfo{journal}{Appl. Sci.}} \textbf{\bibinfo{volume}{9}},
  \bibinfo{pages}{1169} (\bibinfo{year}{2019}).

\bibitem{shashkin2021metal}
\bibinfo{author}{Shashkin, A.~A.} \& \bibinfo{author}{Kravchenko, S.~V.}
\newblock \bibinfo{title}{Metal–insulator transition and low-density phases
  in a strongly-interacting two-dimensional electron system}.
\newblock \emph{\bibinfo{journal}{Ann. Phys.}} \textbf{\bibinfo{volume}{435}},
  \bibinfo{pages}{168542} (\bibinfo{year}{2021}).

\bibitem{Hykin}
\bibinfo{author}{Haykin, S.~O.}
\newblock \emph{\bibinfo{title}{Neural Networks: A Comprehensive Foundation,
  2nd Edition}} (\bibinfo{publisher}{Pearson}, \bibinfo{year}{1999}).

\bibitem{Zupan}
\bibinfo{author}{Zupan, J.} \& \bibinfo{author}{Gasteiger, J.}
\newblock \emph{\bibinfo{title}{Neural Networks for Chemists: An Introduction
  1st Edition}} (\bibinfo{publisher}{VCH Publishers}, \bibinfo{year}{1999}).

\bibitem{kohonen2001}
\bibinfo{author}{Kohonen, T.}
\newblock \emph{\bibinfo{title}{Self-Organizing Maps}}
  (\bibinfo{publisher}{Springer, Berlin, Heidelberg}, \bibinfo{year}{2001}).

\bibitem{Vesanto1999}
\bibinfo{author}{Vesanto, J.}
\newblock \bibinfo{title}{{SOM-based data visualization methods}}.
\newblock \emph{\bibinfo{journal}{Intell. Data Anal.}}
  \textbf{\bibinfo{volume}{3}}, \bibinfo{pages}{111--126}
  (\bibinfo{year}{1999}).

\bibitem{heemskerk98}
\bibinfo{author}{Heemskerk, R.} \& \bibinfo{author}{Klapwijk, T.~M.}
\newblock \bibinfo{title}{Nonlinear resistivity at the metal-insulator
  transition in a two-dimensional electron gas}.
\newblock \emph{\bibinfo{journal}{Phys. Rev. B}} \textbf{\bibinfo{volume}{58}},
  \bibinfo{pages}{R1754--R1757} (\bibinfo{year}{1998}).

\bibitem{blatter94}
\bibinfo{author}{Blatter, G.}, \bibinfo{author}{Feigel'man, M.~Y.},
  \bibinfo{author}{Geshkenbein, Y.~B.}, \bibinfo{author}{Larkin, A.~I.} \&
  \bibinfo{author}{Vinokur, V.~M.}
\newblock \bibinfo{title}{Vortices in high-temperature superconductors}.
\newblock \emph{\bibinfo{journal}{Rev.\ Mod.\ Phys.}}
  \textbf{\bibinfo{volume}{66}}, \bibinfo{pages}{1125--1388}
  (\bibinfo{year}{1994}).

\bibitem{Res1}
\bibinfo{author}{Zhou, Z.} \emph{et~al.}
\newblock \bibinfo{title}{Resistance and current-voltage characteristics of
  individual superconducting {Nb}{Se}$_{2}$ nanowires}.
\newblock \emph{\bibinfo{journal}{Phys. Rev. B}} \textbf{\bibinfo{volume}{76}},
  \bibinfo{pages}{104511} (\bibinfo{year}{2007}).

\bibitem{Res2}
\bibinfo{author}{Buchacek, M.} \emph{et~al.}
\newblock \bibinfo{title}{Experimental test of strong pinning and creep in
  current-voltage characteristics of type-{II} superconductors}.
\newblock \emph{\bibinfo{journal}{Phys. Rev. B}}
  \textbf{\bibinfo{volume}{100}}, \bibinfo{pages}{224502}
  (\bibinfo{year}{2019}).

\bibitem{Res3}
\bibinfo{author}{Jensen, H.~J.}, \bibinfo{author}{Brass, A.},
  \bibinfo{author}{Brechet, Y.} \& \bibinfo{author}{Berlinsky, A.~J.}
\newblock \bibinfo{title}{Current-voltage characteristics in a two-dimensional
  model for flux flow in type-{II} superconductors}.
\newblock \emph{\bibinfo{journal}{Phys. Rev. B}} \textbf{\bibinfo{volume}{38}},
  \bibinfo{pages}{9235--9237} (\bibinfo{year}{1988}).

\bibitem{Res4}
\bibinfo{author}{Willa, R.}, \bibinfo{author}{Koshelev, A.~E.},
  \bibinfo{author}{Sadovskyy, I.~A.} \& \bibinfo{author}{Glatz, A.}
\newblock \bibinfo{title}{Strong-pinning regimes by spherical inclusions in
  anisotropic type-{II} superconductors}.
\newblock \emph{\bibinfo{journal}{Supercond. Sci. Technol.}}
  \textbf{\bibinfo{volume}{31}}, \bibinfo{pages}{014001}
  (\bibinfo{year}{2018}).

\bibitem{dolgopolov2018similarity}
\bibinfo{author}{Dolgopolov, V.~T.}
\newblock \bibinfo{howpublished}{personal communication}
  (\bibinfo{year}{2018}).

\bibitem{weHF}
\bibinfo{author}{Kagalovsky, V.}, \bibinfo{author}{Kravchenko, S.} \&
  \bibinfo{author}{Nemirovsky, D.}
\newblock \bibinfo{title}{Hartree-{F}ock description of a {W}igner crystal in
  two dimensions}.
\newblock \emph{\bibinfo{journal}{Physica E: Low-Dimens. Syst. Nanostruct.}}
  \textbf{\bibinfo{volume}{119}}, \bibinfo{pages}{114016}
  (\bibinfo{year}{2020}).

\end{thebibliography}

\end{document}